# A PRECISION MEASUREMENT OF THE CASIMIR FORCE FROM 0.1 TO 0.9 μm

*U. Mohideen\*, and Anushree Roy.*
*Dept. of Physics, Univ. of California, Riverside, CA 92521.*

## ABSTRACT

We have used an atomic force microscope to make precision measurements of the Casimir force between a metallized sphere of diameter 196 μm and flat plate. The force was measured for plate-sphere surface separations from 0.1 to 0.9 μm. The experimental results are consistent with present theoretical calculations including the finite conductivity, roughness and temperature corrections. The root mean square average deviation of 1.6 pN between theory and experiment corresponds to a 1% deviation at the smallest separation.

In 1948 H.B.G. Casimir, calculated an extraordinary property that two uncharged metallic plates would have an attractive force in vacuum [1]. This results from an alteration by the metal boundaries of the zero point electromagnetic energy, that pervades all of space as predicted by quantum field theory[1-3]. Similar forces result when the strong or gravitational forces are altered by boundaries [3,4]. In the case of the strong force examples include atomic nuclei which confine quarks and gluons [3]. Because of the topological dependence of the Casimir force, the nature and value of this force can also imply a choice between a closed or open universe and the number of space-time dimensions [3,4]. Here we report a precision measurement of the Casimir force between a metallized sphere of diameter 196 µm and a flat plate using an Atomic Force Microscope (AFM). The measurement is consistent with corrections calculated to date. Given the broad implications of the Casimir force, precision measurements would motivate the development of accurate theories on the mechanical forces resulting from zero point energy density[5].

Initially the Casimir force was thought to be similar to the van der Waals force which is an attractive force between two neutral molecules [2]. The van der Waals force results from the fluctuating dipole moment of the materials involved. Lifshitz [6] generalized the van der Waals force between two extended bodies as the force between fluctuating dipoles induced by the zero point electromagnetic fields. The Lifshitz theory [6] and the related Casimir-Polder force [7] have been experimentally verified with reasonable agreement to the theory [8,9]. However, it was soon realized that unlike the van der Waals force, the Casimir force is a strong function of geometry and that between two halves of thin metal spherical shells is repulsive [2-4,10]! Despite the enormous theoretical activity (please see Ref. [3]), there have been only two experimental attempts at observing the Casimir force [11,12]. The first by Sparnaay in 1958 [11] was not conclusive due to 100% uncertainty in the measurements. Last year, in a landmark experiment [12] using a torsion pendulum, Lamoureaux clearly demonstrated the presence of the Casimir force. Though the reported statistical precision was ±5%, significant corrections (>20%) due to the finite conductivity of the metal surface were not observed [12]. Also the roughness correction [13,14] was not observed or estimated. This was probably due to the large experimental systematic error (the electrostatic force between surfaces was 5 times the Casimir force), or due to a fortuitous cancellation of all corrections[13]. Nevertheless, the experiment has been used to set important theoretical constraints[15]. Thus there is a strong need to improve the experimental precision and check the validity of the theoretical corrections.

The Casimir force for two perfectly conducting parallel plates of area '$A$' separated by a distance '$d$' is: $F(d) = -\frac{\pi^2 \hbar c}{240} \frac{A}{d^4}$. It is strong function of '$d$' and is measurable only for $d < 1$ μm. Experimentally it is hard to configure two parallel plates uniformly separated by distances less than a micron. So the preference is to replace one of the plates by a metal sphere of radius '$R$' where $R \gg d$. For such a geometry the Casimir force is modified to [12,16]:

$$F_c^0(d) = \frac{-\pi^3}{360} R \frac{\hbar c}{d^3} \quad . \tag{1}$$

As the surfaces are expected to form a boundary to the electromagnetic waves there is a correction due to the finite conductivity of the metal. This correction to second order based on the free electron model of the reflectivity of metals [13,17] for a given metal plasmon frequency $\omega_p$ is:

$$F_c^p(d) = F_c^0(d)\left[1 - 4\frac{c}{d\omega_p} + \frac{72}{5}\left(\frac{c}{d\omega_p}\right)^2\right] \quad . \tag{2}$$

Given the small separations '$d$' there are also corrections to the Casimir force resulting from the roughness of the surface given by [13,14]:

$$F_c^R(d) = F_c^p(d)\left[1 + 6\left(\frac{A_r}{d}\right)^2\right], \tag{3}$$

where '$A_r$' is the average roughness amplitude and equal roughness for both surfaces has been assumed. There are also corrections due to the finite temperature [12,18] given by:

$$F_c(d) = F_c^R(d)\left(1 + \frac{720}{\pi^2} f(\xi)\right) \quad , \tag{4}$$

where $f(\xi) = (\xi^3/2\pi)\zeta(3) - (\xi^4\pi^2/45)$, $\xi = 2\pi k_B T d/hc = 0.131 \times 10^{-3} d$ nm$^{-1}$ for T= 300°K, $\zeta(3)=1.202…$ is the Riemann zeta function and $k_B$ is the Boltzmann constant.

We use a standard AFM to measure the force between a metallized sphere and flat plate at a pressure of 50 mTorr and at room temperature. A schematic diagram of the experiment is shown in figure 1. Polystyrene spheres of 200±4 µm diameter were mounted on the tip of 300 μm long cantilevers with Ag epoxy. A 1.25 cm diameter optically polished sapphire disk is used as the plate. The cantilever (with sphere) and plate were then coated with 300 nm of Al in an evaporator. Aluminum is used because of its high reflectivity for wavelengths (sphere-plate separations) >100nm and good representation of its reflectivity in terms of a plasma wavelength $\lambda_p$~100nm [19]. Both surfaces are then coated with less than 20 nm layer of 60%Au/40% Pd (measured >90% transparency for $\lambda$<300nm [20]). This was necessary to prevent any space charge effects due to patch oxidation of the Al coating. A Scanning Electron Microscope (SEM) image of the coated cantilever with sphere attached is shown in Figure 2. The sphere diameter was measured using the SEM to be 196 μm. The average roughness amplitude of the metallized surfaces was measured using an AFM to be 35 nm.

In the AFM, the force on a cantilever is measured by the deflection of its tip. A laser beam is reflected off the cantilever tip to measure its deflection. A force on the sphere would result in a cantilever deflection leading to a difference signal between photodiodes A and B (shown in Fig. 1). This force and the corresponding cantilever deflection are related by Hooke's Law: F=k $\Delta z$, where 'k' is the force constant and '$\Delta z$' is the cantilever deflection. The piezo extension with applied voltage was calibrated with height standards and its hysteresis was measured. The corrections due to the piezo hysteresis (2% linear correction) and cantilever deflection (to be discussed later) were applied to the sphere-plate separations in all collected data.

To measure the Casimir force between the sphere and plate they are grounded together with the AFM. The plate is then moved towards the sphere in 3.6 nm steps and the corresponding photodiode difference signal was measured (approach curve). The signal obtained for a typical scan is shown in Figure 3a. Here '0' separation stands for contact of the sphere and plate surfaces. It does not take into account the absolute average separation $\geq 120$ nm due to the 20 nm Au/Pd layer (transparent at these separations[20]) and the 35 nm roughness of the Al coating on each surface. Region 1, shows that the force curve at large separations is dominated by a linear signal. This is due to increased coupling of scattered light into the diodes from the approaching flat surface. Embedded in the signal is a long range attractive electrostatic force, from the contact potential difference between the sphere and plate, and the Casimir force (small at such large distances). In region-2 (absolute separations between contact and 350 nm) the Casimir force is the dominant characteristic far exceeding all the systematic errors (the electrostatic force is less than 3% of the Casimir force in this region). Region-3 is the flexing of the cantilever resulting from the continued extension of the piezo after contact of the two surfaces. Given the distance moved by the flat plate (x-axis), the difference signal of the photodiodes can be calibrated to a cantilever deflection in nanometers using the slope of the curve in region-3. The deflection of the cantilever leads to a decrease in the sphere-plate separation in Region 1&2 which can be corrected by use of the slope in Region-3. This cantilever deflection correction to the surface separation is of order 1% and is given as: $d = d_{piezo} - F_{pd}/m$, where 'd' is the corrected separation between the two surfaces, '$d_{piezo}$' is the separation from the voltage applied to the piezo i.e x-axis of figure 3a, 'm' is the slope of the linear curve in Region-3 and '$F_{pd}$' is the photodiode difference signal shown along the y-axis in figure 3a. The use of Hooke's law to describe the force is validated by the linearity of the photodiode difference signal with cantilever deflection in Region-3.

Next, the force constant of the cantilever was calibrated by an electrostatic measurement. The sphere was grounded to the AFM and different voltages in the range of $\pm 0.5$ V to $\pm 3$ V were applied to the plate. The force between a charged sphere and plate is given as[21]:

$$F = 2\pi\varepsilon_o (V_1 - V_2)^2 \sum_{n=1}^{\infty} \operatorname{csch} n\alpha (\coth\alpha - n\coth n\alpha) \quad . \quad (5)$$

Here '$V_1$' is the applied voltage on the plate and '$V_2$' represents the residual charge on the grounded sphere. $\alpha = \cosh^{-1}(1+d/R)$, where $R$ is the radius of the sphere and '$d$' is the separation between the sphere and the plate. From the difference in force for voltages $\pm V_1$ applied to the plate, we can measure the residual potential on the grounded sphere '$V_2$' as 29 mV. This residual potential is a contact potential that arises from the different materials used to ground the sphere. The electrostatic force measurement is repeated at 5 different separations and for 8 different voltages '$V_1$'. Using Hooke's law and the force from eq. 5 we measure the force constant of the cantilever 'k'. The average of all the measured 'k' was 0.0182 N/m.

The systematic error corrections to the force curve of figure 3a, due to the residual potential on the sphere and the true separations between the two surfaces, are calculated similar to Ref. [12]. Here the near linear force curve in region-1, is fit to a function of the form: $F = F_c(d+d_o) + B/(d+d_o) + C \times (d+d_o) + E$. Here '$d_o$' is the absolute separation on contact, which is constrained to 120±5 nm, is the only unknown to be completely obtained by the fit. The second term represents the inverse linear dependence of the electrostatic force between the sphere and plate for $R>>d$ as given by eq. 5 (verified during the force calibration step)[12]. $B = -2.8$ nN-nm corresponding to $V_2 = 29$ mV and $V_1 = 0$ in eq. 5 is used. The third term represents the linearly increasing coupling of the scattered light into the photodiodes and E is the offset of the curve. Both $C$ and $E$ can be estimated from the force curve at large separations. The best fit values of $C$, $E$ and the absolute separation '$d_o$' are determined by minimizing the $\chi^2$. The finite conductivity correction and roughness correction (the largest corrections) do not play a significant role in region-1 and thus the value of '$d_o$' determined by the fitting is unbiased with respect to these corrections. These values of $C$, $E$ and $d_o$ are then used to subtract the systematic errors from the force curve in region-1 and 2 to obtain the measured Casimir force as: $F_{c-m} = F_m - B/d - C \times d - E$ where $F_m$ is the measured force. Figure 3b is the measured Casimir force corresponding to the force curve of figure 3a. The solid line is the theoretical Casimir force curve of eq.4 with the finite conductivity, roughness and temperature corrections.

This procedure is repeated for 26 scans in different locations of the flat plate. The average measured Casimir force $F_{c-m}$ as a function of sphere-plate separation from all the scans is shown in figure 4 as solid squares. The height of the solid square corresponds to the average deviation observed in the 26 scans. The theoretical Casimir force from eq. 4, with no adjustable parameters, is shown as a solid line. The plasma frequency '$\omega_p$' corresponding to a wavelength of 100 nm was used in the theoretical curve [19]. The root mean square (rms) deviation $\sigma = \sqrt{\frac{(F_{expt} - F_{th})^2}{N}} = 1.6 \text{pN}$, where $F_{expt}$ and $F_{th}$ are the experimental and theoretical Casimir force values respectively and '$N=256$' is the number of data points. This deviation $\sigma$ which is 1% at the smallest surface separation, can be taken as a statistical measure of the experimental precision. The dash-dot line is the Casimir force without the finite conductivity, roughness or temperature corrections (eq.1) which results in a $\sigma$ of 6.3 pN (5% deviation at the smallest surface separation) between experiment and theory. The dashed line includes only the finite conductivity correction (eq.2) which results in a $\sigma$ of 5.5 pN (5% deviation at the smallest separation). The dotted line includes only the roughness correction leading to a $\sigma$ of 48 pN (40% deviation at the closest separation). The theoretical finite conductivity correction and roughness correction are observed to be consistent with the measurement. The experiment has been repeated for different cantilevers, spheres and plates.

In conclusion, we have performed a precision measurement of the Casimir force between a metallized sphere and flat plate using an AFM. The measured Casimir force is consistent with the corrections for the finite conductivity and roughness of the metal surfaces. With lithographic fabrication of cantilevers with large radius of curvature, interferometric detection of cantilever deflection and use of lower temperatures to reduce thermal noise a factor of over 1000 improvement in the precision should be possible in future using this technique. Given the broad implications of the Casimir force such precision measurements should allow for careful checks of the mechanical properties of vacuum.

[*] To whom correspondence should be addressed. Email address: umar.mohideen@ucr.edu

**FIGURE CAPTIONS**

Figure 1: Schematic diagram of the experimental setup. Application of voltage to the piezo results in the movement of plate towards the sphere. The experiments were done at a pressure of 50 mTorr and at room temperature.

Figure 2: Scanning Electron Microscope image of the metallized sphere mounted on a AFM cantilever.

Figure 3: (a) A typical force curve as a function of the distance moved by the plate, (b) the measured Casimir force corresponding to (a) as a function of sphere-plate surface separation. The solid line is the theoretical Casimir force from eq. 4.

Figure 4: The measured average Casimir force as a function of plate-sphere separation for 26 scans is shown as square dots. The error bars show the range of experimental data at representative points. The theoretical Casimir force from eq. 4 with all corrections, is shown as a solid line. The rms deviation between the experiment and theory is 1.6 pN. The dash-dot line is the Casimir force without the finite conductivity, roughness or temperature correction (eq.1) which results in a rms deviation of 6.3 pN. The dashed line includes only the finite conductivity correction (eq.2) which results in a rms deviation of 5.5 pN. The dotted line includes only the roughness correction leading to a rms deviation of 48 pN.

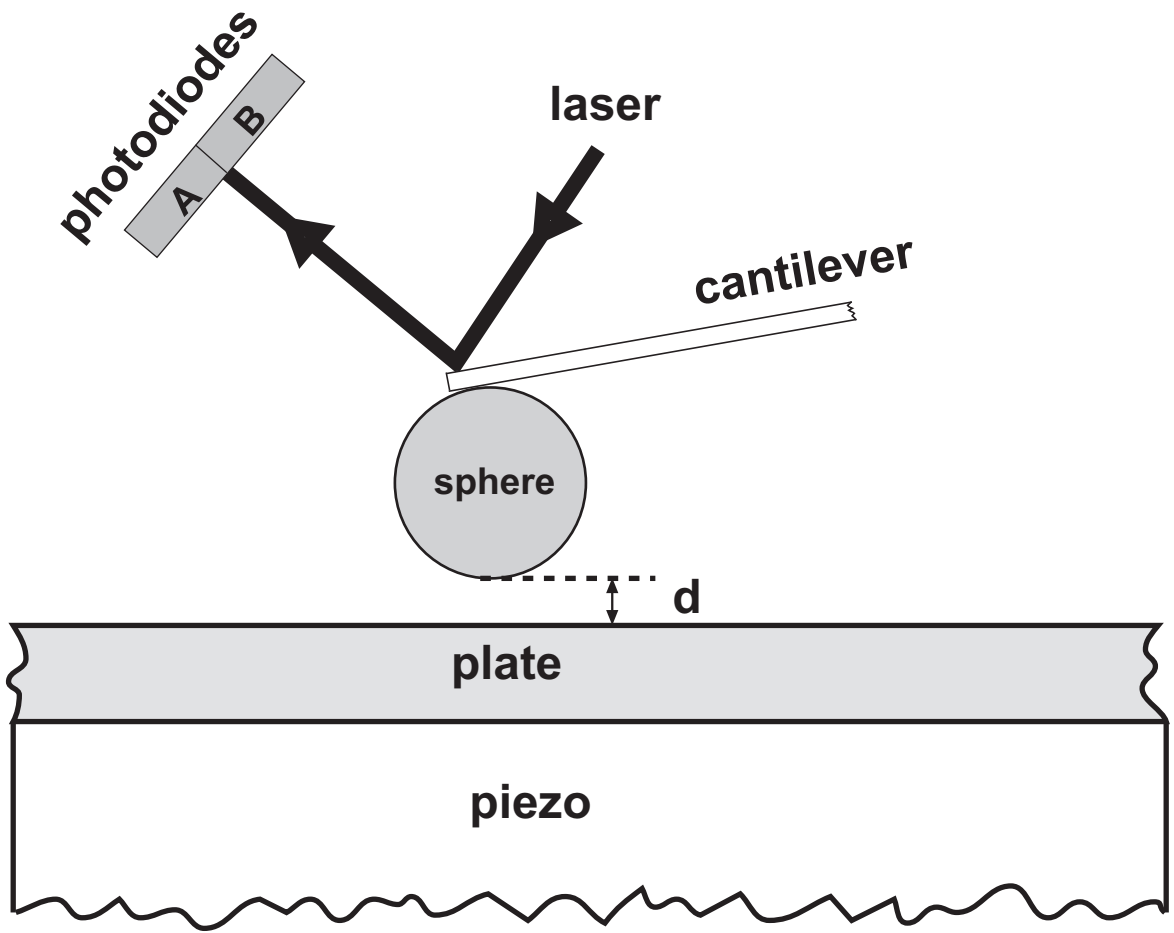